\documentclass[aps, prb, superscriptaddress, reprint, showpacs]{revtex4-1}

\usepackage{amsfonts}
\usepackage{amsmath}
\usepackage{amssymb, bm}
\usepackage{graphicx, epstopdf, color}
\usepackage{soul}

\begin{document}

\title{Nonlocal transistor based on pure crossed Andreev reflection in a EuO-graphene/superconductor hybrid structure}

\author{Yee Sin Ang}
\affiliation{Engineering Product Development, Singapore University of Technology and Design, Singapore 487372}
\affiliation{SUTD-MIT International Design Center, Singapore University of Technology and Design, Singapore 487372}
\affiliation{School of Physics, University of Wollongong, Wollongong, NSW 2522, Australia}

\author{L. K. Ang}
\affiliation{Engineering Product Development, Singapore University of Technology and Design, Singapore 487372}
\affiliation{SUTD-MIT International Design Center, Singapore University of Technology and Design, Singapore 487372}
\author{C. Zhang}
\affiliation{School of Physics, University of Wollongong, Wollongong, NSW 2522, Australia}
\author{Zhongshui Ma}
\affiliation{School of Physics, Peking University, Beijing 100871, China}
\affiliation{Collaborative Innovation Center of Quantum Matter, Beijing 100871, China}

\begin{abstract}

We study the interband transport in a superconducting device composed of graphene with EuO-induced exchange interaction. We show that pure crossed Andreev reflection can be generated exclusively without the parasitic local Andreev reflection and elastic cotunnelling over a wide range of bias and Fermi levels in an EuO-graphene/superconductor/EuO-graphene device. The pure non-local conductance exhibits rapid on/off switching and oscillatory behavior when the Fermi levels in the normal and the superconducting leads are varied. The oscillation reflects the quasiparticle propagation in the superconducting lead and can be used as a tool to probe the subgap quasiparticle mode in superconducting graphene, which is inaccessible from the current-voltage characteristics. Our results suggest that the device can be used as a highly tunable transistor that operates purely in the non-local and spin-polarized transport regime.


\end{abstract}

\maketitle

\emph{Introduction} - Andreev reflection (AR) is the excitation of a hole in a normal/superconductor interface when two opposite-spin electrons are coupled into a Cooper pair in the superconductor \cite{andreev}. In a normal/superconductor/normal (N/S/N) three-terminal geometry, two electrons can couple locally in the same normal lead or non-locally in different normal leads to form a Cooper pair in the superconductor. The local coupling produces the `usual' Andreev reflection (AR) while the non-local coupling produces the exotic \emph{crossed Andreev reflection} (CAR) \cite{byers}. The reverse process of CAR in N/S/N device has been proposed as the basis of Cooper pair splitter that generates entangled electron pair in condensed matter environment \cite{samuelsson, samuelsson2, prada, lesovik, schroer}. High Cooper pair splitting efficiency of $\sim 90\%$ has been experimentally achieved in a carbon nanotube-based N/S/N device \cite{schindele}. Moreover, the pairing symmetry of a superconductor can also be probed by CAR signal \cite{benjamin}. Unfortunately, the generation of CAR-dominated transport is challenging since it is inevitably plagued by electron elastic co-tunnelling (EC) and local AR \cite{falci}.

\begin{figure}[t]
\includegraphics[width = 6.75cm]{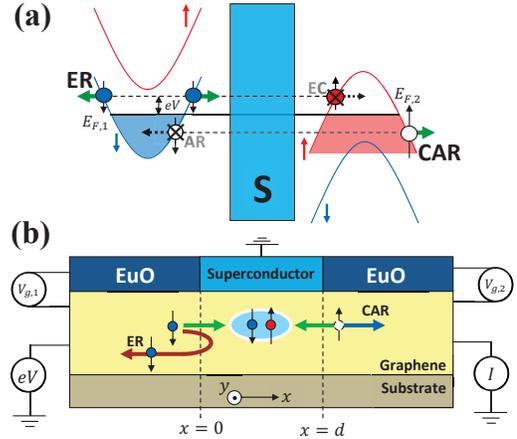}
\caption{(a) Mechanism of pCAR in a gapped and spin-split dispersion; (b) Schematic of the EuO-G/superconductor/EuO-G device. The incident energy is related to the bias by $E = eV$.}
\label{fig:schematic}
\end{figure}

Generating \emph{pure CAR} (pCAR) using energy band topology was first proposed in a graphene bipolar transistor \cite{cayssol}. By precisely tuning the Fermi levels and the bias voltage, EC and local AR excitations are forced to lie exactly on the Dirac points. Due to the vanishing quasiparticle density, EC and AR are completely eliminated. Despite its conceptual simplicity, the experimental realization is difficult since precisely fixed Fermi levels and bias are required. A significant improvement can be achieved by using a gapped energy dispersion \cite{veldhorst}. AR and EC are completely blocked by the whole \emph{continuum} of the bandgap instead of a single Dirac point, thus lifting the constraint on the bias voltage. pCAR mediated by \emph{bandgap blocking} can occur in semiconductor \cite{veldhorst}, silicene \cite{linder2}, MoS$_2$ \cite{bai} and quantum spin hall insulator \cite{chen}, provided that the Fermi levels are placed within one superconducting gap with respect to the conduction and valence band extrema.

The stringent condition of having precisely fixed Fermi levels can be circumvented, for example, by engineering the valley-helicity \cite{cresti} of zigzag graphene nanoribbon \cite{wang} and by shifting the valley-spin splitting \cite{xiao} in the valence band of MoS$_2$ \cite{majidi}. In systems with tunable bandgap such as silicene and bilayer graphene \cite{drummond, zhang_etc}, the existence of 1D topologically protected edge state \cite{martin} provides another opportunity to create widely tunable pCAR. Remarkably, the suppressed intervalley scattering forces a further removal of the normal electron reflection (ER) \cite{wang_J}. Beyond supercoductivity, tunable pCAR has been predicted \cite{veldhorst2} in the topological exciton condensate in 3D topological insulator \cite{seradjeh}. This offers an exciting alternative condensed matter platform to generate entangled electrons.

\emph{Theoretical concept} - We propose a different strategy to achieve widely tunable pCAR in this work. We show that the \emph{interband transport} in a gapped and spin-split energy dispersion can sustain pCAR over a wide range of bias and Fermi levels. The concept is illustrated in Fig. \ref{fig:schematic}(a). Consider the case where the Fermi level of the incident side, $E_{F,1}$, is placed between the two \emph{conduction} spin-subband edges and that of the transmitted side, $E_{F,2}$, is placed between the two \emph{valence} spin-subband edges. For an incident electron residing in the lower conduction spin-subband, no \emph{opposite-spin} electrons are available below the Fermi level for local AR. In the transmitted region, spin conservation forbids the electron from tunnelling into the opposite-spin valence subband. As a result, the only permissible processes are ER and the much sought-after pCAR. The conditions of having precisely tuned bias and Fermi levels are \emph{both} relaxed. To demonstrate this, we consider an Europium oxide-graphene (EuO-G) ferromagnetic hybrid-structure \cite{forster} [\ref{fig:schematic}(b)]. First-principle calculations predicted that EuO strongly spin-polarizes the $\pi$-orbitals of graphene \cite{yang, su} and induces a large exchange splitting. A sizable spin-dependent bandgap, which crucially blocks the local AR and EC excitations, is present. We found that the non-local conductance in EuO-G/S/EuO-G exhibits fast on-off switching via normal leads gating. Furthermore, the non-local conductance exhibits an oscillatory behavior with the superconductor-gate that directly reflects the subgap superconducting Dirac quasiparticle propagation. We found that a minimal subthreshold swing of $15.1$ mV and a large on-off ratio of $10^5$ can be achieved. This opens up the possibility of high efficiency graphene-based \emph{non-local transistor} in which all local and non-entangled processes are suppressed.  

\begin{figure}
\includegraphics[width = 8.5cm]{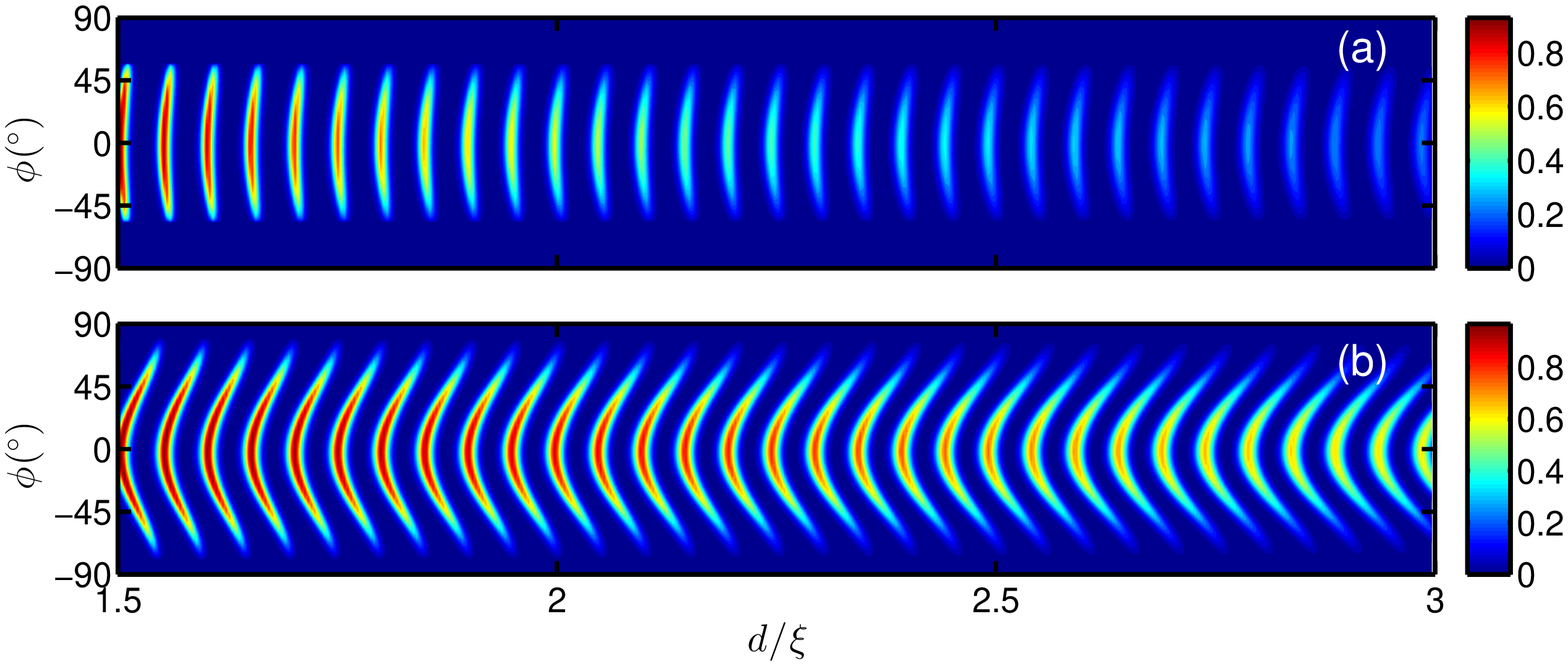}
\includegraphics[width = 8.5cm]{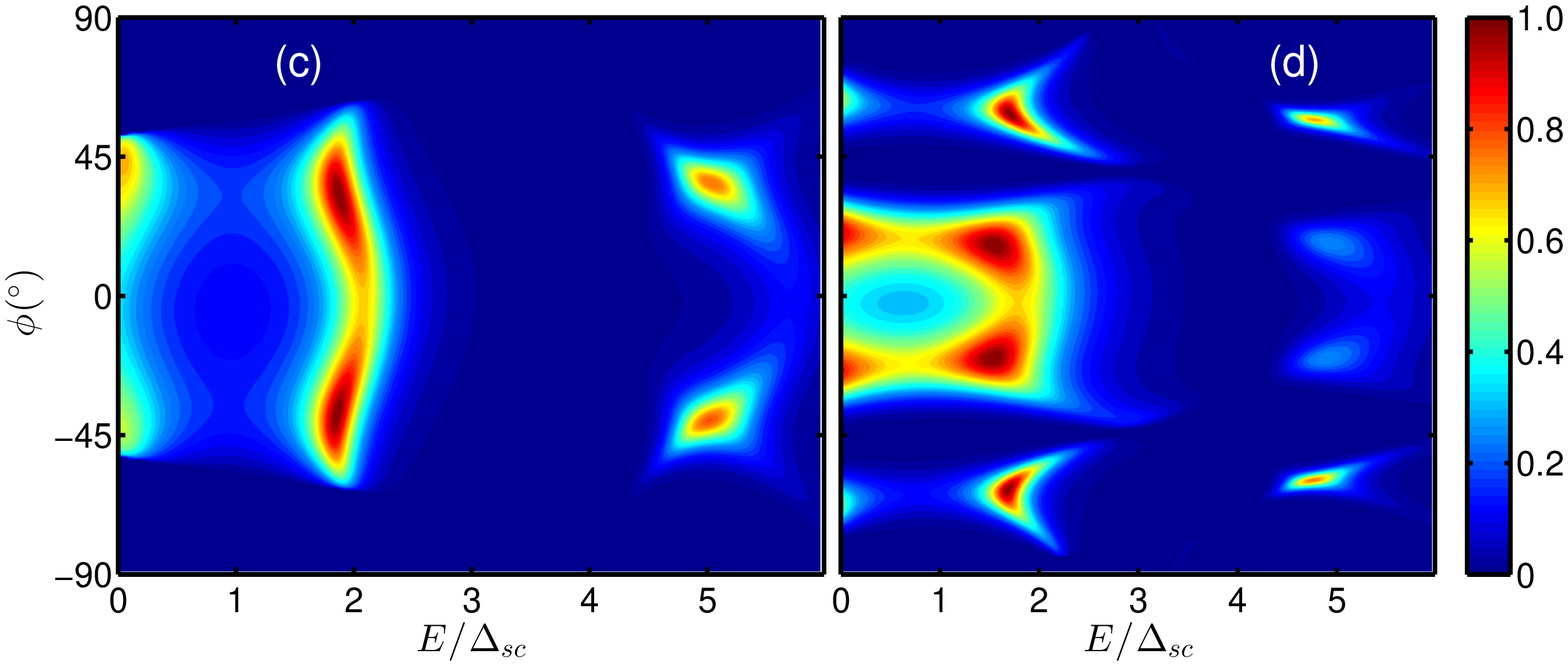}
\caption{$T_{CAR}$ as a function of superconductor width $d$ for (a) $(E_{F,1}, E_{F,2}) = (30, -40)$ meV; and (b) $(E_{F,1}, E_{F,2}) = (60, -60)$ meV with $E / \Delta_{sc} = 0.9$ ($\xi = \hbar v_F / \pi \Delta_{sc}$). (c) and (d) shows the $T_{CAR}$ as a function of incident energy $E/\Delta_{sc}$ at the same Fermi levels as (a) and (b), respectively, with $d = 2.5\xi$ ($\Delta_{sc} = 1$ meV and $\mu_S = 200$ meV).}
\label{fig:tcar}
\end{figure}

\emph{Model} - In EuO/G, the $K$ and $K'$ Dirac cones are mapped onto the $\Gamma$ point due to the Brillouin zone folding \cite{yang, su}. The low energy effective Hamiltonian can be written as \cite{song} $\mathcal{H}_{\bm{k},\sigma} = \sigma h \mathcal{I} + \Delta_{\sigma} \tau_z + \hbar v_\sigma \bm{k} \cdot \bm{\tau}$, where $\sigma = \pm 1$ for spin-up and spin-down electrons, $h$ is the proximity-induced exchange interaction, $\bm{k} = (k_x, k_y)$ is the electron wavevector, $\bm{\tau} = (\tau_x, \tau_y, \tau_z)$ are the Pauli matrices, $v_{\sigma}$ and $\Delta_{\sigma}$ are the spin-dependent Fermi velocity and bandgap, respectively. $\mathcal{I}$ is a $2\times2$ identity matrix. The eigenenergy is $\varepsilon_{\sigma\eta}(\bm{k}) = \eta \sqrt{\Delta_\sigma^2 + \hbar v_\sigma^2k^2} + \sigma h$ where $k = |\bm{k}|$ and $\eta = \pm 1$ denotes conduction and valence bands. The normalized eigenstate is $\xi_{\sigma\eta}(\bm{k}) = \left[\left(\varepsilon_{\sigma\eta}(\bm{k}) - \Delta_\sigma \right) / 2 \varepsilon_{\sigma\eta}(\bm{k}) \right]^{1/2} \left[\frac{\hbar v_\sigma k_x - ik_y}{\varepsilon_{\sigma\eta}(\bm{k}) - \Delta_\sigma}, 1 \right]^T$, where $T$ stands for transpose. First-principle calculation \cite{yang} gives $v_\sigma = (1.4825 - 0.1455\sigma) \times v_F$, $ h = 31$ meV and $\Delta_\sigma = (58 + 9\sigma)$ meV. The Bogoliubov-de Gene (BdG) equation \cite{BdGenne} is given as
\begin{equation}\label{eq:bdg}
\begin{pmatrix}
\mathcal{H}_{\bm{k},\sigma}(x) - E_F \mathcal{I} & \Delta_{sc}(x)\mathcal{I} \\
\Delta_{sc}^*(x) \mathcal{I}&  -\left(\mathcal{H}_{\bm{k},\bar{\sigma}}(x) - E_F \mathcal{I}\right)  
\end{pmatrix}
\begin{pmatrix}
u_{\sigma} \\
v_{\bar{\sigma}}
\end{pmatrix}
= 
E_{\sigma} (\mathbf{k})
\begin{pmatrix}
u_{\sigma} \\
v_{\bar{\sigma}}
\end{pmatrix}
\end{equation}
where $\bar{\sigma} = -\sigma$, $h(x) = h$ for $0>x>d$ and the superconducting gap is $\Delta_{sc}(x) = \Delta_{sc}$ for $0<x<d$. We take the phase in $\Delta_{sc}$ as zero. For $\Delta_{sc}(x)=0$, Eq. (\ref{eq:bdg}) can be decoupled into a spin-$\sigma$ electron part and a spin-$\bar{\sigma}$ hole part. As $k_y$ is a good quantum number, we write $k_\pm = -i \partial/\partial x \pm ik_y$ and solve Eq. (\ref{eq:bdg}) for the quasiparticle eigenstates and the excitation energies. The transport coefficients can then be straightforwardly obtained by matching the wavefunctions at $x=0$ and at $x=d$ \cite{supp}.

\emph{Results \& discussions} -  In the numerical calculation, we choose $\Delta_{sc} = 1$ meV which agrees with recent experimental value \cite{efetov}. For conciseness, we focus on the pCAR transport phenomenon originating from an incident electron residing in the $\sigma = -1$ conduction subband and transmitted as a purely $\sigma = +1$ polarized valence hole. Since there is a large common-gap of $(\Delta_{+} + \Delta_{-} -2h) = 54$ meV, only conduction hole is involved in the quasiparticle transport \cite{beenakker}. According to first-principle results \cite{yang, song}, the Fermi levels lie in the ranges of 18 meV $ < E_{F,1} < 98$ meV and $- 80$ meV $< E_{F,2} < -36$ meV. This corresponds to a wide windows of $\Delta E_F^{(c)} = 80$ meV and $\Delta E_F^{(v)} = 44$ meV for conduction and valence bands respectively. We first study the pCAR transmission probabilities, $T_{CAR}$, in Fig. \ref{fig:tcar}. The angle of incidence of the electron is denoted by $\phi$. $T_{CAR}$ oscillates rapidly with $d$ because of the quasiparticle interference in the superconducting gap [Fig. \ref{fig:tcar}(a)]; transmission peaks occurs whenever the subgap superconducting quasiparticle wavevector matches the resonance wavevector $k_0 = n\pi /d$. A significant difference between Figs. \ref{fig:tcar}(a) and \ref{fig:tcar}(b) is that the resonance `stripes' are almost vertical and well-separated in Fig. \ref{fig:tcar}(a) (Fermi levels lie closer to the band edges) while in Fig. \ref{fig:tcar}(b) (Fermi levels lie farther away from the band edges) the resonance patterns are curved and are no longer well-separated. This contrasting behavior can be seen in the $eV$-dependence of $T_{CAR}$. Along a vertical cut at $E/\Delta_{sc}=2$ and $E/\Delta_{sc} = 5$, four $T_{CAR}$ hotspots are clearly present in Fig. \ref{fig:tcar}(d) instead of only two in Fig. \ref{fig:tcar}(c). The four-hotspot is caused by the curved resonance pattern in Fig. \ref{fig:tcar}(b) as it is composed of two pairs of transmission resonance at constant $d/\xi$: one from the central region of a resonance `stripe' and one from the tail region of the preceding curved resonance `stripe'.

\begin{figure}
\includegraphics[width = 8.5cm]{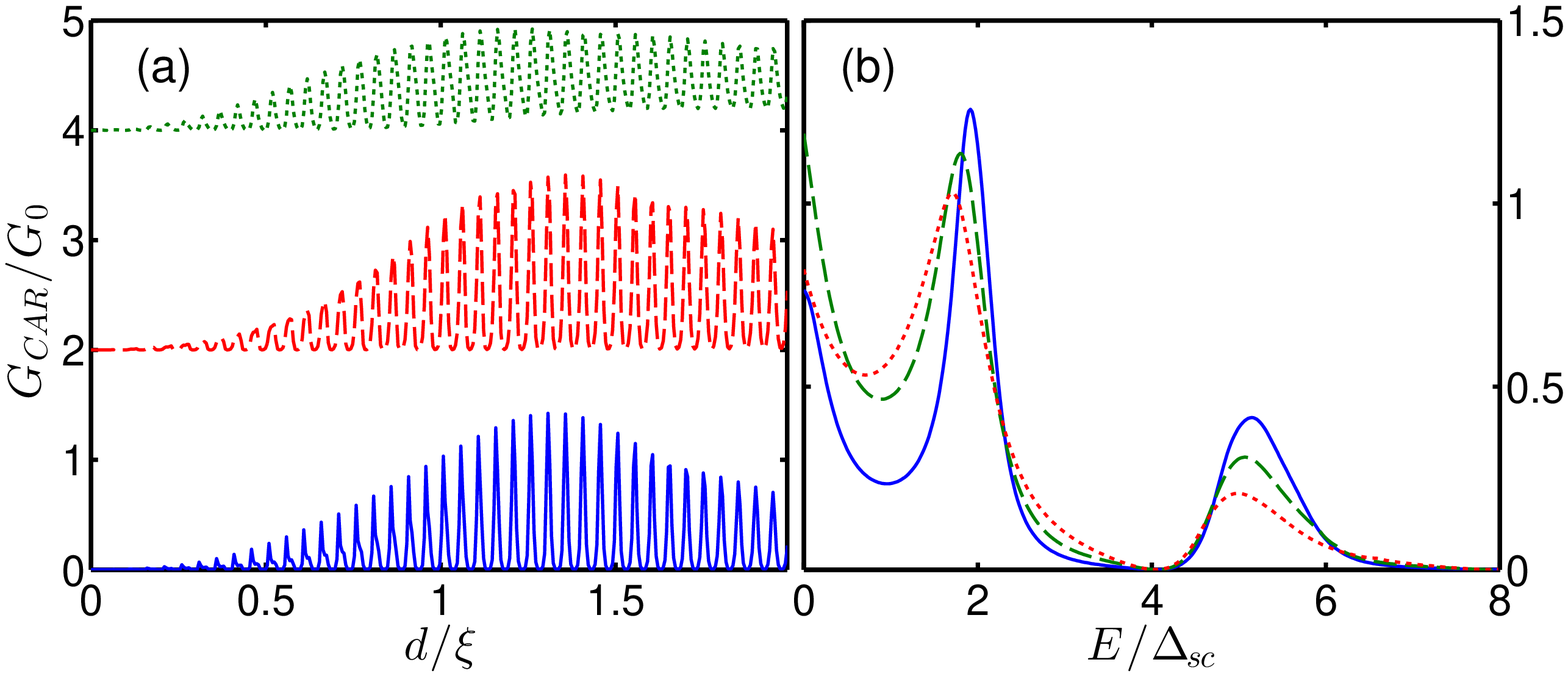}
\includegraphics[width = 8.5cm]{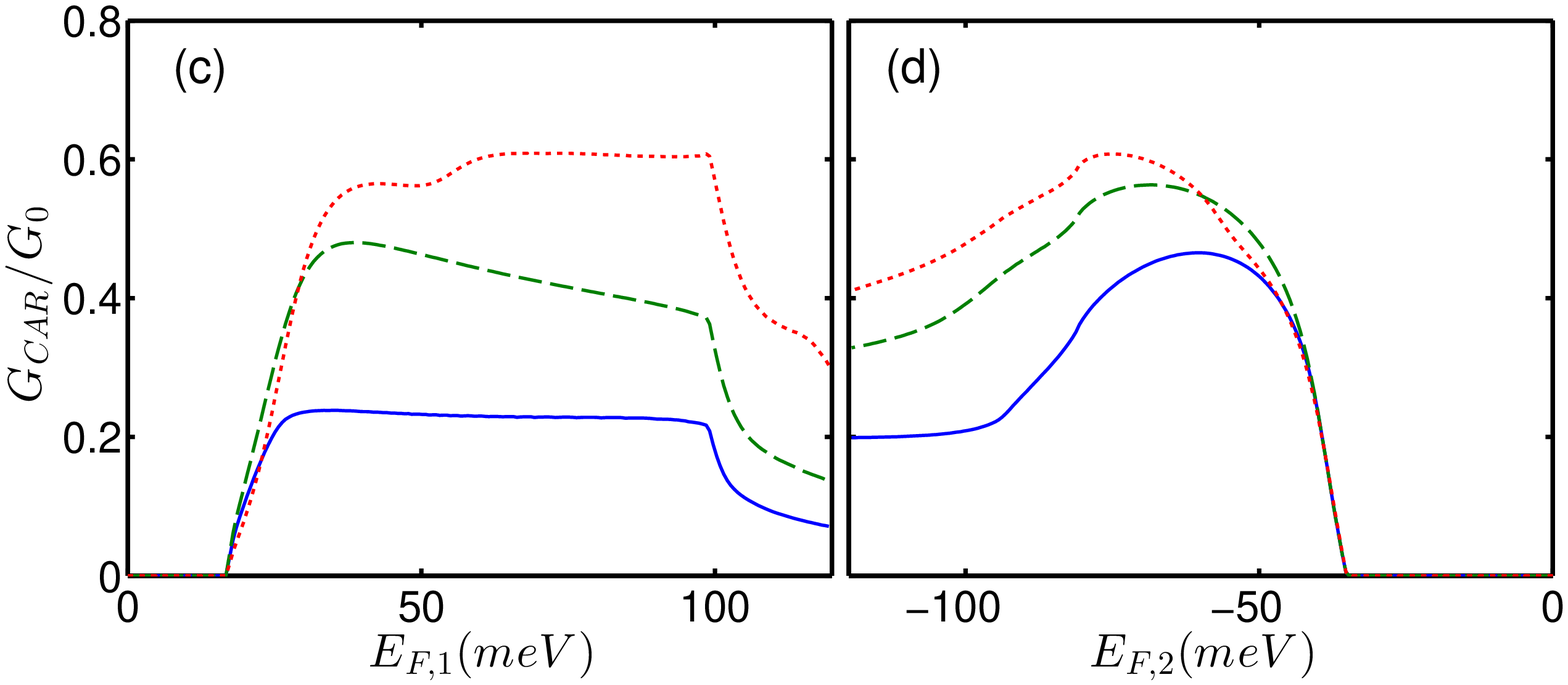}
\caption{Non-local conductance $G_{CAR}$. (a) $d$-dependence for (solid) $(E_{F,1}, E_{F,2}) = (30, -40)$ meV, (dashed) $(E_{F,1}, E_{F,2}) = (30, -60)$ meV and (dotted) $(E_{F,1}, E_{F,2}) = (60, -60)$ meV at $E/\Delta_{sc} = 0.9$ (Data are offset vertically by $2G_0$ for clarity and $E=0.9\Delta_{sc}$); (b) $E/\Delta_{sc}$-dependence with the same Fermi levels as (a) and $d = 2.5 \xi$; (c) $E_{F,1}$-dependence with $E_{F,2} = -40, -50, -70$ meV (solid, dashed and dotted line, respectively); and (d) $E_{F,2}$ dependence with $E_{F,1} = 30, 40, 60$ meV (solid, broken and dotted line, respectively). $d = 2.5 \xi$ and $E/\Delta_{sc} = 0.9$for (c) and (d). When $E_{F,1}>98$ meV and $E_{F,2}< -80$ meV, $G_{CAR}$ decreases significantly due to the onset of the competing local AR and EC processes.}
\label{fig:cond}
\end{figure}

The zero-temperature \emph{non-local conductance} generated by pCAR is given as \cite{btk} $G_{CAR}/G_0 = \int T_{CAR} \cos{\phi} \textrm{d}\phi$. $G_0$ is the ballistic normal conductance in $\sigma = -1$ channel. In Fig. \ref{fig:cond}(a), $G_{CAR}$ as a function of the superconductor width, $d$, is plotted. $G_{CAR}$ oscillates rapidly with $d$ due to the fast oscillation of $T_{CAR}$. Interestingly, $G_{CAR}$ minima are near-zero only when the $E_F$'s are close to the band edge. This is a direct consequence of the well-separated resonance `stripes' as discussed in Fig. \ref{fig:tcar}(a). In Fig. \ref{fig:cond}(b), the $G_{CAR}$-resonances occur at $E/\Delta_{sc} \approx 2$ and $\approx 5$. This is consistent with the $T_{CAR}$ hotspots observed in Figs. \ref{fig:tcar}(c) and (d). Note that although the $T_{CAR}$ peaks originates from the Fabry-P\'erot interference (FB) inside $\Delta_{sc}$, tunnelling current of solely pCAR is unachievable via FB alone since $G_{CAR}$ is an \emph{angular-averaged} quantity. The selective enhancement of one transport process and the simultaneous suppression of the rest is only achievable at certain incident angles for a given energy. Without filtering out the local AR and EC processes via the band topology, the tunnelling conductance is inevitably mixed with local and non-entangled components. In the proposed device, $(E_{F,1}, E_{F,2})$ can be tuned without destroying the pCAR. $G_{CAR}$ as a function of $(E_{F,1},E_{F,2})$ is calculated in Figs. \ref{fig:cond}(c) and \ref{fig:cond}(d). The onset of $G_{CAR}$ for incident energy $E$ is $(18 - E)$ meV and $(-36 + E)$ meV for $E_{F,1}$ and $E_{F,2}$, respectively. Before these onsets, $G_{CAR}$ is completely switched-off due to the depletion of the charge carriers. Remarkably, $G_{CAR}$ rises very sharply post onset, suggesting a potential in fast on-off switching application. To estimate the switching characteristic, we define the \emph{Fermi level subthreshold swing} as $SS(E_{F,i}) = \left( \text{d} log_{10} I_{CAR} / \text{d} E_{F,i} \right)^{-1} \approx \left(\Delta log_{10} G_{CAR} / \Delta E_{F,i}\right)^{-1}$ where $i=1,2$ denotes the two normal leads. We found that in the linear-growth regime immediately after the onset, $SS(E_{F,1})$ is about 7.1 meV/dec (meV per decade). For $E_{F,2}$, the onset of $G_{CAR}$ is even sharper, yielding $SS(E_{F,2})\approx 3.3$ meV/dec. The \emph{gate-voltage subthreshold swing}, $SS(V_{g,i}) = \left( \text{d} log_{10} I_{CAR} / \text{d} V_{g,i} \right)^{-1}$, can be estimated from experimental data \cite{gel-chen, shi}. We found that \cite{gate-voltage} $SS(V_{g,1}) \approx 60.5$ mV/dec and $SS(V_{g,2}) \approx 15.1$ mV/dec. The remarkably small $SS(V_{g,2})$ shows an even steeper on-off switching characteristic in comparison with state-of-the-art MoS$_2$-based transistor recently reported in \cite{sarkar}. This reveals the potential of the proposed device as a fast switching transistor that operates uniquely in the non-local and $100\%$ spin-polarized transport regime. 

\begin{figure}
\includegraphics[width = 8.5cm]{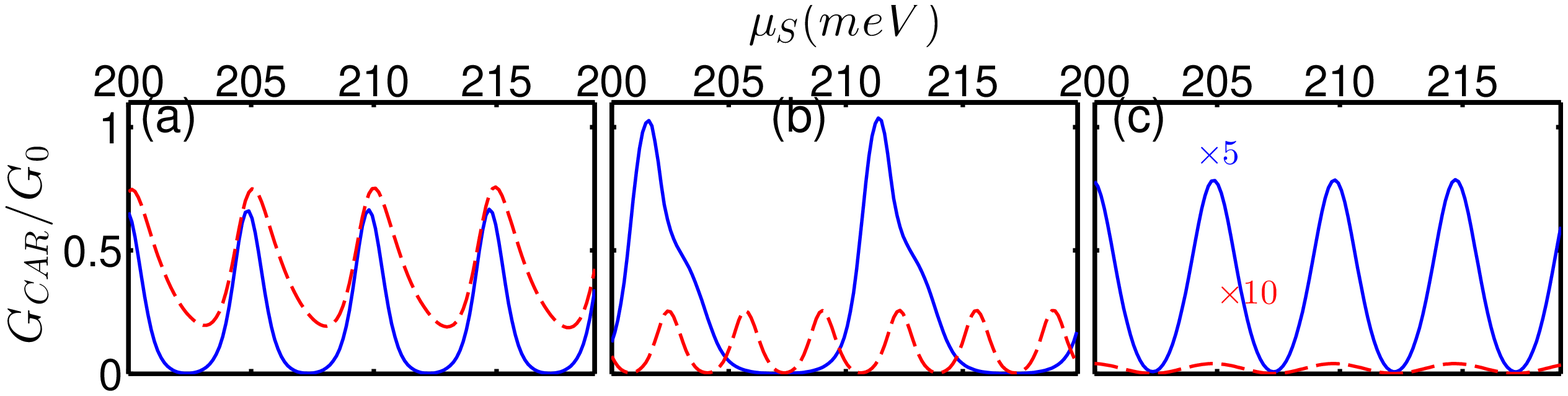}
\includegraphics[width = 8.5cm]{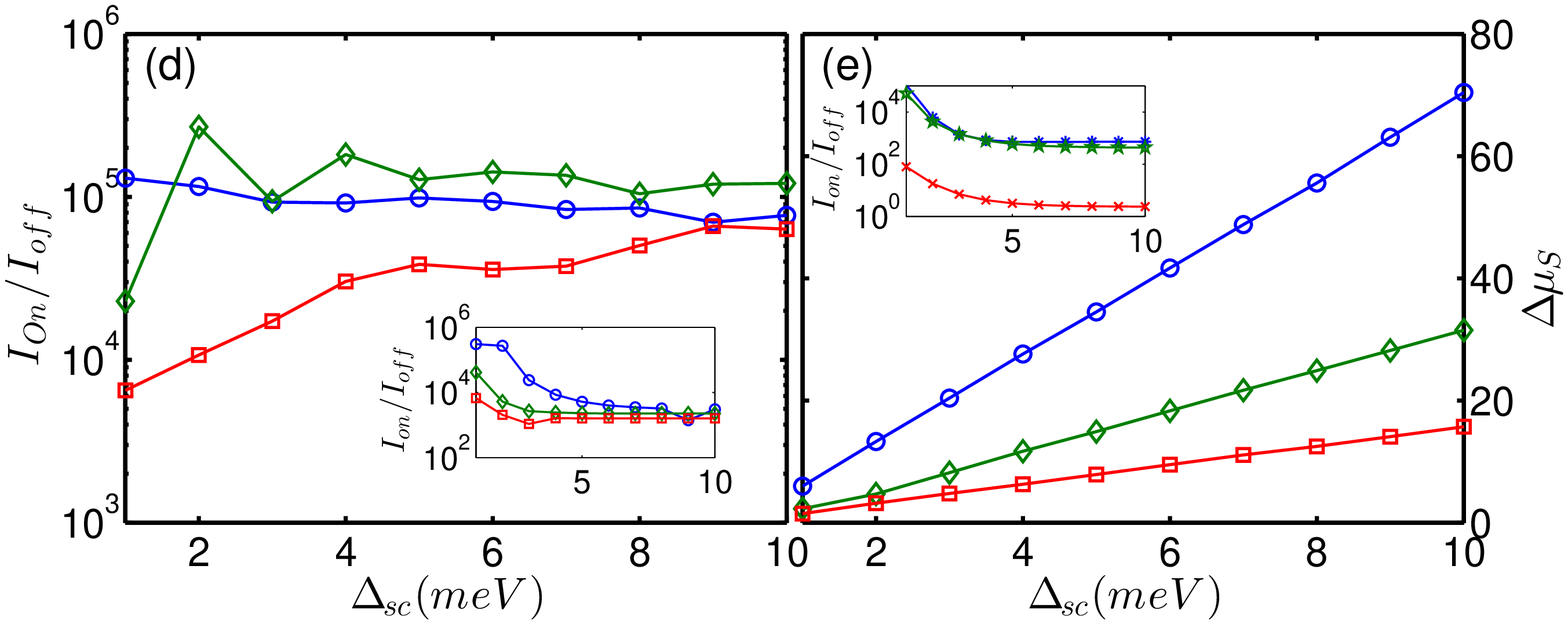}
\caption{$\mu_S$ dependence of $G_{CAR}$ with $E/\Delta_{sc} = 0.9$. (a) Fermi levels $(E_{F,1}, E_{F,2})$ are $(30, -40)$ meV (solid) and $(60, -60)$ meV at $d = 2\xi$ (dashed); (b) $d = \xi$ (solid) and $d = 3\xi$ (dashed); (c) $\Delta_{sc} = 2$ meV (solid) and $\Delta_{sc} = 5$ meV (dashed) with $d_0$ fixed at $\hbar v_F/\tilde{\Delta}_{sc}$ where $\tilde{\Delta}_{sc} = 1$ meV. In (b) and (c), $(E_{F,1}, E_{F,2}) = (30, -40)$ meV are used. The $\Delta_{sc}$-dependence of (d) $I_{\text{on}} / I_{\text{off}}$ and (e) $\Delta \mu_s$ with $E/\Delta_{sc} = 0.9$ and $(E_{F,1}, E_{F,2}) = (18, -36)$ meV. The widths are $d=\xi$ $(\circ)$, $d=2\xi$ $(\diamond)$  and $d=3\xi$ $(\square)$. Inset in (d) is the same as the main plot except that $d=d_0$ $(\circ)$, $d=2d_0$ $(\diamond)$  and $d=3d_0$ $(\square)$. Inset in (e) shows the $I_{\text{on}} / I_{\text{off}}$ at several sets of $(\Delta_{+}, \Delta_{-}, h)$, i.e. $(170, 150, 90)$ meV ($\ast$), $(210, 190, 10)$ meV ($\star$) and $(30, 10, 2)$ meV ($\times$). $E/\Delta_{sc} = 0.9$, $d = d_0$ and the Fermi levels are fixed at the band edges.}
\label{fig: mu_s}
\end{figure}

We calculated $G_{CAR}$ as a function of the superconducting graphene Fermi level, $\mu_S$, in Figs. \ref{fig: mu_s}(a) to (c). In general, $G_{CAR}$ exhibits oscillatory behavior with $\mu_S$. The followings are observed: (i) the oscillation frequency is unaffected by $(E_{F,1}, E_{F,2})$ [Fig. \ref{fig: mu_s}(a)]; (ii) the frequency of $G_{CAR}$ oscillation is reduced by a smaller $d$ [Fig. \ref{fig: mu_s}(b)]; and (iii) the oscillation frequency is unaffected by $\Delta_{sc}$ but the amplitude is severely damped at larger $\Delta_{sc}$ [Fig. \ref{fig: mu_s}(c)]. These oscillatory behaviors reflect the subgap ($E<\Delta_{sc}$) quasiparticle dynamics residing in the superconducting graphene. For $E<\Delta_{sc}$, the superconducting Dirac quasiparticle wavevector is composed of a propagating (real) term, $k_S = \sqrt{(\hbar v_F)^{-2}\mu_S^2 +  q^2}$, and an imaginary (damping) term, $\kappa = \left(\Delta_{sc}/\hbar v_F k_S\right) \sin\left( \cos^{-1}\left(E/\Delta_{sc}\right) \right)$ \cite{beenakker}. $\mu_S$-tuning directly modifies $k_S$. When $k_S$ is tuned across two successive subgap standing-wave modes, a peak-valley-peak $G_{CAR}$-oscillation is produced. $(E_{F,1}, E_{F,2})$ do not play a direct role in $k_S$. Hence, they do not alter the oscillation frequency [Fig. \ref{fig: mu_s}(a)]. When $d$ is decreased, the difference between two successive standing-wavevectors becomes larger as the quantized standing-wavevector $k_0 \propto 1/d$. The peak-to-peak transition thus requires a larger range of $\mu_S$ to be scanned across. This results in a reduced oscillation frequency as seen in Fig. \ref{fig: mu_s}(b). $\Delta_{sc}$ affects only the damping term as $\kappa \propto \Delta_{sc}$ when $E/\Delta_{sc} \to 1$. Increasing $\Delta_{sc}$ thus leads to a stronger damping which reduces the amplitude without changing its oscillation frequency. Physically, one can interpret the $\Delta_{sc}$-dependence as followed: a larger $\Delta_{sc}$ leads to a shorter coherence length $\xi \propto 1/\Delta_{sc}$. At a fixed $d_0$, the `effective' barrier width becomes larger in the relative sense of $d_0/\xi$. Therefore, the CAR tunnelling current is heavily damped as the non-local Cooper pairing of electrons has to overcome too many coherence lengths.

The $G_{CAR}$-oscillation offers an additional tunable parameter to control the transport by gating the superconducting graphene. As $\mu_S$-tuning cannot completely switch the $G_{CAR}$ off and $G_{CAR}$ oscillates periodically, we define two quantities to characterize the $\mu_S$-switching effect: (i) the non-local current on-off ratio, $I_{\text{on}} / I_{\text{off}} \approx G_{CAR}^{(\text{max})} / G_{CAR}^{(\text{min})}$ for small bias where $G_{CAR}^{(\text{max})}$ and $G_{CAR}^{(\text{min})}$ are the maximum and minimum conductance determined at the vicinity of $\mu_S \approx 200$ meV, respectively; and (ii) the range of $\mu_S$ required for peak-to-valley switching, $\Delta \mu_S$. For small $d$, $I_{\text{on}} / I_{\text{off}}$ can be as high as $10^5$ over a wide range of $\Delta_{sc}$ [Fig. \ref{fig: mu_s}(d)]. Large $\Delta \mu_S$ is desirable for efficient $\mu_S$-switching so that the valley-to-peak transition is robust against the Fermi level fluctuation induced by charge inhomogeneity and substrate \cite{xue}. The $\Delta \mu_S$ is in the undesirably small values of few meV at small $\Delta_{sc}$ due to the rapid $G_{CAR}$-oscillation. Interestingly, $\Delta\mu_S$ increases linearly with $\Delta_{sc}$ [Fig. \ref{fig: mu_s}(e)] and can be improved to $70$ meV at $\Delta_{sc} = 10$ meV. The linear relation between $\Delta\mu_S$ and $\Delta_{sc}$ can be explained by noting that $d$ is in the unit of $\xi\propto 1/\Delta_{sc}$ and hence is not fixed in Figs. \ref{fig: mu_s}(d) and (e). Since the oscillation frequency is determined by the quantized standing-wavevector which is $k_0 \propto d$ and $d\propto 1/\Delta_{sc}$, we have $k_0 \propto \Delta_{sc}$, which leads to the linear dependence. In the insets, we calculated $I_{\text{on}}/I_{\text{off}}$ as a function of $\Delta_{sc}$ with $d$ in the unit of fixed unit $d_0 = \hbar v_F / \pi \Delta_{0}$ where $\Delta_0 = 1$ meV [Fig. \ref{fig: mu_s}(d)] and for various $\Delta_{\sigma}$ and $h$ at $d = d_0$ [Fig. \ref{fig: mu_s}(e)]. We observe that $I_{\text{on}} / I_{\text{off}}$ is significantly reduced at small $\Delta_{\sigma}$ and at large $\Delta_{sc}$. This confirms the importance of a large spin-dependent bandgap and finite $h$ in achieving efficient non-local current gating. Furthermore, strong Cooper pairing does not lead to enhanced non-local transport.

\emph{Conclusion} - In conclusion, we proposed widely tunable pCAR in the \emph{interband} transport of spin-split and gapped dispersion in EuO-G/S/EuO-G. The proposed device exhibits rapid on/off switching which can potentially be used as a building block in CAR-based quantum computing and spintronics. We emphasize that the pCAR mechanism proposed here is fundamentally different from the case of MoS$_2$ with exchange interaction \cite{majidi}. In our scheme, pCAR is based on the \emph{interband} quasiparticle transport between the conduction and the valence spin-split subbands via Fermi levels tuning. Due to the \emph{inverted band topology} between conduction and valence spin-split subbands, the elimination of the local AR and EC branches can be straightforwardly achieved without the need to shift the relative separation between the subbands via exchange interaction \cite{majidi}. For EuO-G/S single-interface, local AR is completely suppressed in the regime studied here. As it is well-known that local AR generates Joule heating that undesirably lowers the cooling power of a normal/insulator/superconductor-based electronic refrigerator \cite{nahum, bardas, leivo}, we expect EuO-G/S to exhibit enhanced sub-Kelvin cooling performance. One major challenge to observe the rapid $G_{CAR}$-oscillation is the fabrication of high quality sample as the interface roughness and Fermi level fluctuations can wash out the oscillation. Finally, we point out that the pCAR mechanism proposed here is universally applicable to systems with similar band topology such as YiG-graphene \cite{zwang} and monolayer transition-metal dichalcogenides with magnetic doping \cite{cheng} or with proximity to EuO \cite{qi}. These structures offer alternative platforms to test the validity of our prediction. 

We thank Shi-Jun Liang and Kelvin J. A. Ooi for their fruitful discussions. ZSM is thankful for the support of NSFC (11274013) and NBRP of China (2012CB921300). This work was funded by the Singapore Ministry of Education (MOE) T2 grant (T2MOE1401), and the Australian Research Council Discovery Grant (DP140101501).

\end{document}